\documentclass[lettersize,journal]{IEEEtran}
\usepackage{amsmath,amsfonts,amssymb}
\usepackage{array}
\usepackage{textcomp}
\usepackage{color}
\usepackage{stfloats}
\usepackage{url}
\usepackage{verbatim}
\usepackage{caption} 
\usepackage{graphicx}
\usepackage{subfigure}
\usepackage{float} 
\usepackage{cite}
\usepackage{algorithm}
\usepackage{algorithmic}
\makeatletter
\let\NAT@parse\undefined
\makeatother
\usepackage{hyperref}

\makeatother
\usepackage{multirow}
\usepackage{amsmath}
\usepackage{bm}
\usepackage{ntheorem}

\begin{document}

\title{Geometry-Dependent Radiation of Pinching Antennas: Theory, Simulation, and Measurement}

\author{Haoyang Li, Weidong Liu, Zhensheng Chen, \IEEEmembership{Member,~IEEE}, Chaoyun Song, \IEEEmembership{Senior Member,~IEEE}, Gaojie Chen, \IEEEmembership{Senior Member,~IEEE}\vspace{-1em}
\thanks{Haoyang Li, Zhensheng Chen and Gaojie Chen are with the School of Flexible Electronics (SoFE), Sun Yat-sen University, Shenzhen, Guangdong 518107, China (e-mail: hylifj@foxmail.com; chenzhsh27@mail.sysu.edu.cn; chengj235@mail.sysu.edu.cn).
}
\thanks{Weidong Liu is with the School of Telecommunications Engineering, Xidian University, Xi'an 710071, China, and also with the School of Flexible Electronics (SoFE), Sun Yat-sen University, Shenzhen, Guangdong 518107, China (e-mail: wdlui@stu.xidian.edu.cn).
}
\thanks{Chaoyun Song is with the Department of Engineering, King's College London, WC2R 2LS London, U.K. (e-mail: chaoyun.song@kcl.ac.uk).
}
}


\maketitle

\begin{abstract}
Most existing studies achieve beamforming by adjusting the positions of pinching antennas (PAs) and typically model PAs as isotropic radiators. However, under the dielectric scatterer model, the PA radiation pattern depends on its geometry. This letter investigates the radiation patterns of PAs with different geometries through full-wave simulations and measurements, and demonstrates how geometry influences the radiation directivity. In addition, an arc-shaped PA is introduced to enable transmit-direction control in PA systems. A PA system prototype consisting of a dielectric waveguide, waveguide transitions, and a PA element is proposed. Prototype measurements are used to validate the simulations and to characterize the directivity of square and triangular PAs, and the measurement procedure can be applied to obtain radiation patterns for PAs with general geometries. The simulation and measurement results jointly demonstrate that PA geometry is critical in PA systems because it influences the radiation characteristics significantly.
\end{abstract}


\vspace{-0.1cm}
\section{Introduction}
Next-generation (6G) wireless networks are expected to deliver substantially higher capacity and connectivity, yet their performance is increasingly constrained by notably severe free-space path loss and line-of-sight (LoS) blockage, especially at high carrier frequencies \cite{wang2023on}. Existing solutions only partially address this challenge, such as, massive multiple-input multiple-output (MIMO) can provide beamforming gain but cannot fundamentally resolve blockage and large-scale attenuation, while reconfigurable intelligent surfaces (RISs) may suffer from strong additional path loss due to the cascaded link \cite{ding2025flexible}. To overcome these limitations, recent studies have explored flexible antenna systems, such as movable antenna (MA) and fluid antenna (FA), that can reshape the wireless channel more directly. Pinching antenna (PA) systems as one of the flexible antenna techniques, have emerged as a promising architecture specifically targeting large-scale channel effects \cite{ding2025flexible}. PAs employ a dielectric waveguide as a low-loss transmission line and realize wireless transmission by attaching small dielectric perturbations to the waveguide, thereby coupling out a portion of the guided energy into free space, an approach first reported by NTT DOCOMO in 2022 \cite{fukuda2022pinching}. By pre-deploying waveguides and placing PAs close to users, PA systems effectively create a last-meter link, thereby strengthening LoS connectivity, reducing free-space path loss, and mitigating blockage. More recently, PA systems were investigated across a range of wireless scenarios, including integrated sensing and communication (ISAC), energy-efficient transmission, and secure communications \cite{miao2026multi}, \cite{li2025total}, \cite{zhong2025physical}.

Recent works have established some basic theoretical models of PA systems. In \cite{ding2025flexible}, the authors theoretically characterized PA systems and showed that leveraging large-scale PA position reconfigurability can create strong LoS links and deliver clear performance gains over conventional fixed-antenna systems. In \cite{wang2025modeling}, the authors built physics-based PA radiation and coupling models, derived equal-power and proportional-power waveguide power-allocation models, and proposed penalty-based alternating optimization. In \cite{xu2025rate}, the authors studied downlink rate maximization for a basic single-user PA system by optimizing PA locations under coupled path-loss and dual-phase effects, and proposed a two-stage low-complexity algorithm. Overall, prior works mainly emphasize transmission modeling and optimization for PA-enabled links. Although system-level studies have been explored, investigations into the practical radiated-field characteristics of PAs remain insufficient, and many studies adopt idealized assumptions, which lack the rigorous physical grounding and may not accurately capture practical coupling and radiation behavior.


Motivated by the above discussion, to capture the practical radiation behavior of the PA systems, this work proposes a fabricable single-waveguide single-PA system and conducts both simulations and prototype measurements to verify the correctness and feasibility of the proposed design. The following is a brief summary of our main contributions in this letter:
\begin{enumerate}
  \item A dielectric scatterer model based on evanescent field coupling and the volume equivalence principle is used to demonstrate that the radiation field created by the PAs is related to their geometry.
  \item The simulations show that PA geometry has a strong impact on radiation fields. In particular, square and triangular PAs are investigated, showing that the triangular design achieves higher directivity than the square design. In addition, an arc-shaped PA is introduced as a simple and low-cost approach for tuning the transmission direction.
  \item A prototype PA system is fabricated and measured to validate the simulations. The measured radiation patterns show good agreement with the simulated results.
\end{enumerate}


The rest of this paper is organized as follows. Section \ref{section:sysmodel} first reviews location-tuning-based beamforming and then utilizes a dielectric-scatterer-based theory showing that the radiation field produced by the PAs depends on their geometry. In Section \ref{section:simulation}, simulation settings and results are presented to show the radiation characteristics of different PAs. To verify the validity of the simulations, the prototypes and measurements of the proposed PA systems are discussed in Section \ref{section:measurement}. Finally, Conclusions are drawn in Section \ref{section:con}.

\begin{figure} [t!]
  \centering
  \includegraphics[width= 1.87in, height=1.23in]{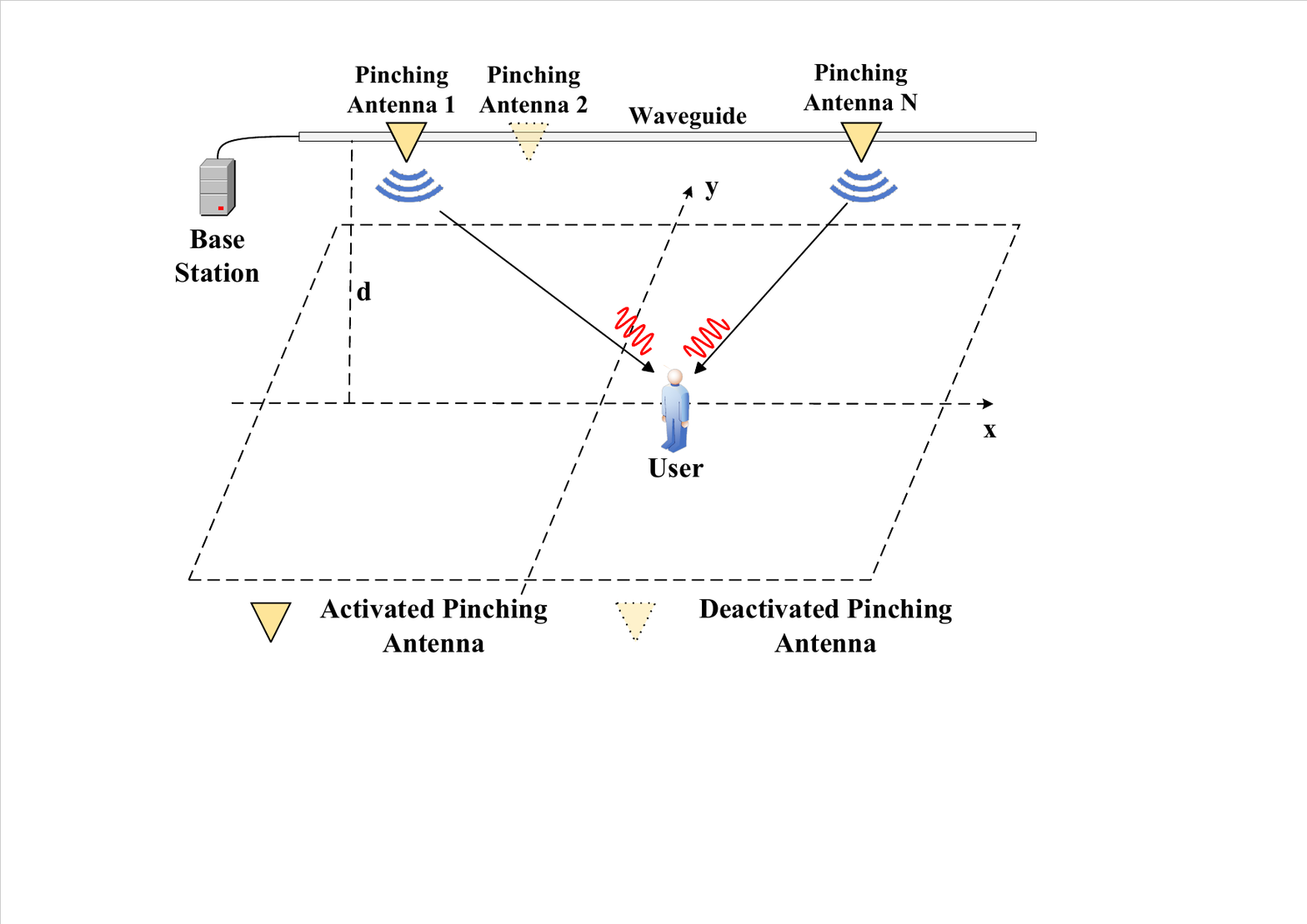}
  \caption{Illustration of a network with a single waveguide and $N$ pinching antennas.}\label{Sys_model.Fig}
\end{figure}

\vspace{-0.1cm}
\section{System Model}\label{section:sysmodel}
PA systems comprise dielectric waveguides integrated with multiple PAs. Fig. \ref{Sys_model.Fig} illustrates a basic PA architecture in which a single waveguide is extended over the $x$-axis at the height $d$ with $N$ PAs. These PAs are arranged sequentially along the waveguide and can be independently activated. According to \cite{ding2025flexible}, \cite{wang2025modeling}, wireless signals can be coupled out of the waveguide when a PA is brought into close proximity to the waveguide or mechanically pinched onto it, such operations are referred to as PA activation.

\subsection{Existing Location Tuning based Beamforming}
Modeling each activated PA as an isotropic radiator, consider a set of $M$ activated PAs located at $L\triangleq \left\{\mathbf{l}_1,...,\mathbf{l}_M\right\} $, $\mathbf{l}_m = (x_m, y_m, z_m)$ denotes the location of the $m$th activated PA in the coordinates. The corresponding channel coefficient between the $m$th activated PA and the user can be written as \cite{wang2025antenna}, \cite{zhao2025pinching}
\setlength\abovedisplayskip{3pt} 
\setlength\belowdisplayskip{3pt}
\begin{equation}\label{pinch_channel.eq}
  \begin{aligned}
    \mathbf{h}_\mathrm{Pin}(\mathbf{L}) = \left[\frac{\eta^{\frac{1}{2}}e^{-j\frac{2\pi}{\lambda}\mathbf{r}_1(\mathbf{l}_1)}}{\mathbf{r}_1(\mathbf{l}_1)},..., \frac{\eta^{\frac{1}{2}}e^{-j\frac{2\pi}{\lambda}\mathbf{r}_M(\mathbf{l}_M)}}{\mathbf{r}_M(\mathbf{l}_M)}\right]^\mathrm{T},
  \end{aligned}
\end{equation}
where $\lambda$ is the free-space wavelength, $\mathbf{r}_i(\mathbf{l}_i), i=1,...,M$ denotes the propagation distance between the user and the $m$th activated PA located at $\mathbf{l}_m$, $\eta = \frac{c^2}{16\pi^2 f_c^2} $ is a constant, $c$ denotes the speed of light, and $f_c$ is the carrier frequency. Then, the received signal at the user can be written as
\setlength\abovedisplayskip{3pt} 
\setlength\belowdisplayskip{3pt}
\begin{equation}\label{received_signal.eq}
  \begin{aligned}
    y_u = \mathbf{h}_\mathrm{Pin}(\mathbf{L})^\mathrm{H}\mathbf{s} + z,
  \end{aligned}
\end{equation}
where $\mathbf{s}$ is the transmitted signal and $z$ denotes the additive white Gaussian noise at user. Assuming the equal power model for the PAs, the transmitted signal can be expressed as
\setlength\abovedisplayskip{3pt} 
\setlength\belowdisplayskip{3pt}
\begin{equation}\label{transmit_signal.eq}
  \begin{aligned}
    \mathbf{s} = \sqrt{\frac{P}{M}}\left[e^{-j\theta_1},...,e^{-j\theta_M}\right]^{\mathrm{T}}s,
  \end{aligned}
\end{equation}
where $P$ is the total transmit power, $s$ is the signal passed onto the waveguide, $\theta_m = 2\pi \frac{\left\lvert \mathbf{l}_{\mathrm{feed}} - \mathbf{l}_m \right\rvert }{\lambda_g}$ denotes the phase shift experienced at the $m$th PA, $\mathbf{l}_{\mathrm{feed}}$ denotes the location of the feed point of the waveguide and $\lambda_g$ denotes the waveguide wavelength in a dielectric waveguide \cite{ding2025flexible}.

Based on \eqref{pinch_channel.eq}, \eqref{received_signal.eq} and \eqref{transmit_signal.eq}, the system can adjust the locations of activated PAs. This adjustment allows control over the associated phase shifts and path losses of the radiated field. With this control, the system is able to steer and focus the radiated field directly toward the intended user. In turn, this realizes location-tuning-based beamforming and effectively improves the quality of the received signal.

\begin{figure} [t!]
  \centering
  \includegraphics[width= 3.24in, height=0.84in]{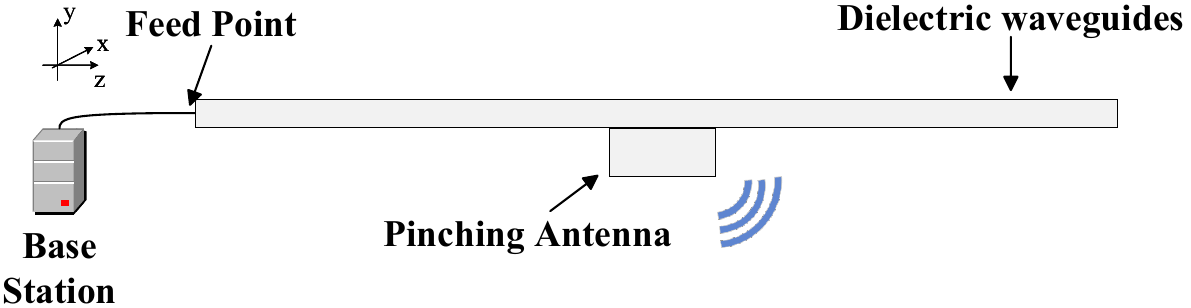}
  \caption{Illustration of the considered pinching antenna system, where the waveguide is equipped with a single pinching antenna.}\label{model.Fig}
\end{figure}

\subsection{Structure-Dependent Radiation of the PA Element}
Most existing studies neglect PA geometry and assume that PAs behave as isotropic radiators. In practice, the PA radiation pattern depends on its geometry. This suggests two key points. First, to fully support existing location-tuning beamforming methods, the PA should be designed to provide near-isotropic radiation. Second, beyond changing PA locations, the transmit direction can also be steered by properly modifying the PA geometry.


To demonstrate the PA radiation features, in this section, we consider a single PA on a single waveguide scenario, as shown in Fig. \ref{model.Fig}. Based on dielectric scatter theory, the excitation and radiation of PA can be described by evanescent field coupling and volume equivalence principle. Considering the dielectric waveguide with an effective refractive index $n_g$ and the transmit signal propagates along the $z$-axis. The electric field distribution of the electromagnetic (EM) wave within the waveguide can be expressed as \cite{okamoto2006fundamentals}
\setlength\abovedisplayskip{3pt} 
\setlength\belowdisplayskip{3pt}
\begin{equation}\label{Ewaveguide.eq}
  \begin{aligned}
    \mathbf{E}_g(x,y,z) = \mathbf{e}_g(x,y)A(z)e^{-j\beta_g z},
  \end{aligned}
\end{equation}
where $\beta_g = \frac{2 \pi n_g}{\lambda}$ is the propagation constant of the waveguide, $A(z)$ represents the modal amplitude along the propagation direction. Due to the boundary conditions, the field components outside the waveguide cross section do not vanish abruptly but instead decay exponentially as an evanescent field. This evanescent field is described by the transverse electric-field distribution function $\mathbf{e}_g(x,y)$.

Outside the waveguide, the field magnitude decays exponentially with the distance $r(x,y)$ away from the waveguide surface as 
\setlength\abovedisplayskip{3pt} 
\setlength\belowdisplayskip{3pt}
\begin{equation}\label{decay.eq}
  \begin{aligned}
    \mathbf{e}_g(r) \propto e^{-\alpha r},
  \end{aligned}
\end{equation}
where $\alpha$ denotes the transverse attenuation constant, which determines the penetration depth of the outward leakage field. The $\alpha$ for a signal with a free-space wavelength $\lambda$ can be expressed as  
\setlength\abovedisplayskip{3pt} 
\setlength\belowdisplayskip{3pt}
\begin{equation}\label{alpha.eq}
  \begin{aligned}
    \alpha = \sqrt{\beta_g^2 - k_0^2} = \frac{2\pi}{\lambda} \sqrt{n_g^2 - 1},
  \end{aligned}
\end{equation}
where $k_0 = 2\pi / \lambda$ is the wavenumber in the free-space. At millimeter-wave (mmWave) frequencies, the free-space wavelength $\lambda$ is small (1~10 mm), so even a slight change in $n_g$ can significantly affect $\alpha$. As a result, the evanescent field is highly sensitive and remains tightly confined to the waveguide surface.

When a dielectric block (i.e., a PA) is placed in close proximity to the waveguide, it introduces a localized dielectric perturbation that modifies the boundary conditions of the guided mode. This perturbation disturbs the modal confinement and enables a portion of the guided energy to couple into the PA through evanescent-field interaction. This coupling excites the intrinsic dipoles of the dielectric PA and generates an equivalent induced polarization current $\mathbf{J}_{eq}$, which can be expressed as
\setlength\abovedisplayskip{3pt} 
\setlength\belowdisplayskip{3pt}
\begin{equation}\label{current.eq}
  \begin{aligned}
    \mathbf{J}_{eq}(\mathbf{r}) = j\omega(\epsilon_\mathrm{PA} - \epsilon_0)\mathbf{E}_\mathrm{inc},
  \end{aligned}
\end{equation}
where $\omega$ denotes the angular frequency, $\epsilon_\mathrm{PA}$ is the permittivity of the PA, $\epsilon_0$ is vacuum permittivity, and $\mathbf{E}_\mathrm{inc}$ denotes the incident electric field that penetrates from the waveguide into the PA.

To illustrate how the excitation source produces free-space radiation, the volume equivalence principle is employed. The far field radiation $\mathbf{E}_\mathrm{far}$ produced by the PA can be regarded as the vector superposition of the contributions from all polarization-current elements within its volume $V$, which can be expressed as 
\setlength\abovedisplayskip{3pt} 
\setlength\belowdisplayskip{3pt}
\begin{equation}\label{e_far.eq}
  \begin{aligned}
    \mathbf{E}_\mathrm{far}(\theta, \phi) = -j\omega\mu_0 \frac{e^{-jk_0r}}{4\pi r} \iiint_V \mathbf{J}_{eq}(\mathbf{r}') e^{jk_0\mathbf{r}' \cdot \mathbf{\hat{r}}} \mathrm{d}V',
  \end{aligned}
\end{equation}
where $\mu_0$ denotes the permeability of free-space, $\mathbf{r}'$ denotes the source-point position vector within the radiating volume, and $\mathbf{\hat{r}}$ denotes the unit vector specifying the far-field observation direction.

Based on \eqref{e_far.eq}, the far-field pattern is determined by the spatial Fourier transform of the induced polarization current distribution over $V$. By reconfiguring the PA geometry, the phase distribution of the induced currents can be reshaped. Consequently, in addition to location-tuning-based beamforming, the PA system can steer the transmission direction through geometry design.

At mmWave frequencies, the PA size is often comparable to the wavelength or even much larger. As a result, the geometric boundary strongly influences the spatial distribution of $\mathbf{J}_{eq}$. The PA shape sets the propagation path length of the excited wave inside the dielectric. Differences in path length create a nonuniform phase profile over the PA surface, which controls the pointing direction of the radiated wavefront. When a user is aligned with the main radiation direction, the received signal is strong. When the user falls in a pattern null, the location-tuning-based beamforming becomes ineffective. 

Based on the above analysis, we obtained the far-field radiation results of several PAs via High Frequecy Struture Simulator (HFSS). Further results will be detailed in next section.

\begin{figure} [t!]
  \centering
  \includegraphics[width= 3.12in, height=0.71in]{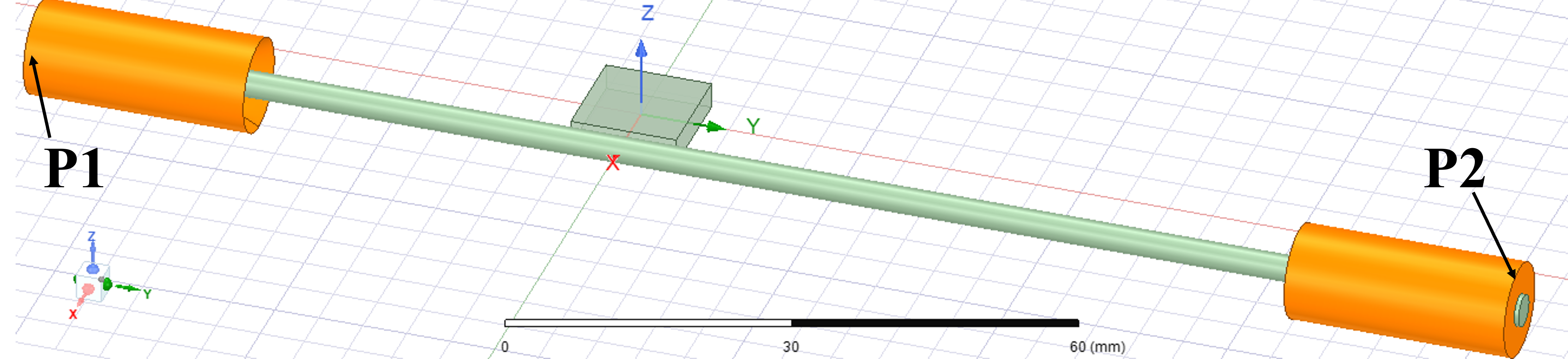}
  \caption{Configuration of the proposed pinching antenna system in simulation with a single PA.}\label{simu_model.Fig}
\end{figure}

\section{Simulation Results}\label{section:simulation}

To illustrate how PA geometry affects the field distribution, this section presents simulations for square and regular triangular PAs. An arc-shaped PA is then introduced to enable flexible control of the radiation direction. 

Fig. \ref{simu_model.Fig} illustrates the proposed PA system configuration used in the simulation. The system consists of a dielectric rod-shaped waveguide, a circular waveguide (CW)-to-dielectric waveguide transition \cite{vu2021experimental}, and a dielectric block functioning as the PA. The signals are fed into the waveguide through one of the waveguide transitions and radiated by the PA. The dielectric waveguide is fabricated from polytetrafluoroethylene (PTFE) with a diameter of 3 mm and a relative permittivity of 2.1. The PA is also made from PTFE, while the waveguide transitions are fabricated from copper. All simulations are analyzed at 60 GHz.

\subsection{Square Pinching Antenna}
\begin{figure}[!t]
  \centering
  \subfigure[]{
  \includegraphics[width=3cm]{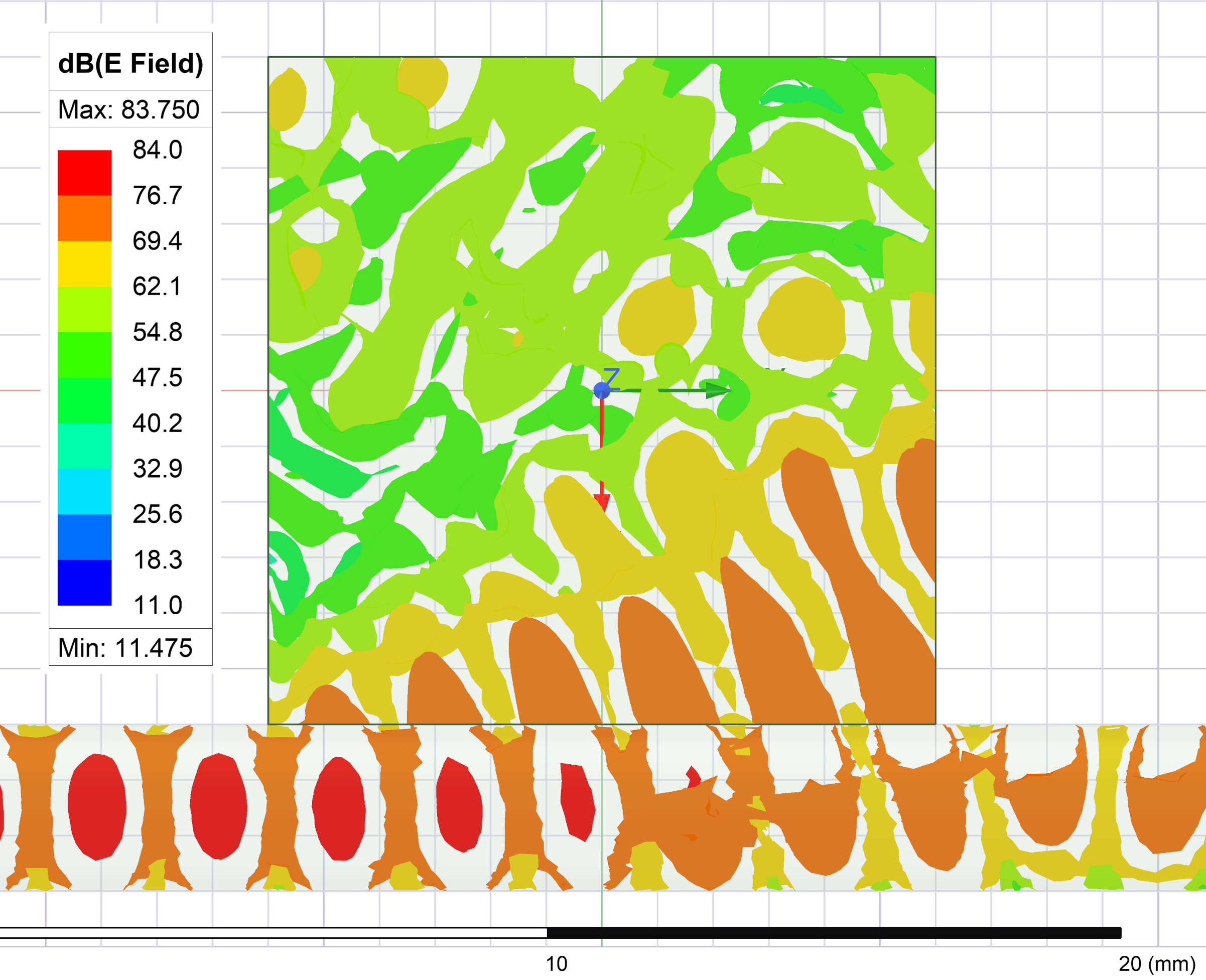}
  }
  \quad
  \subfigure[]{
  \includegraphics[width=3.5cm]{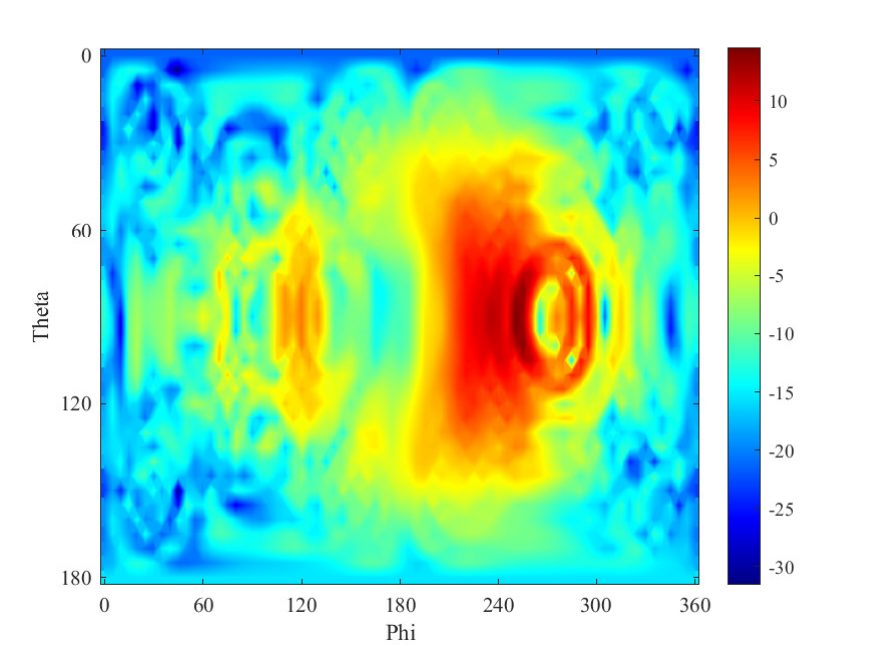}
  }
  \quad
  \subfigure[]{
  \includegraphics[width=3cm]{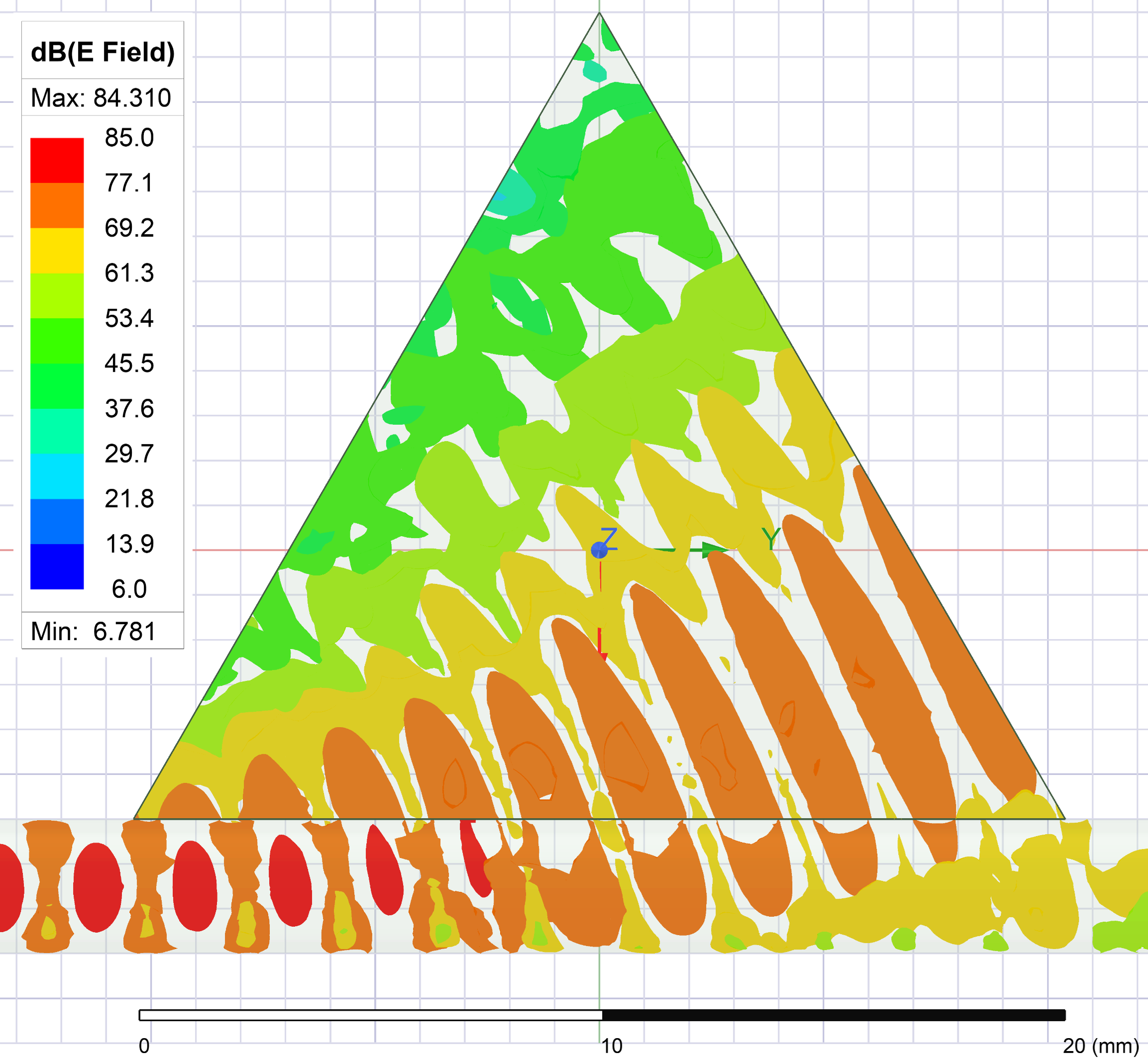}
  }
  \quad
  \subfigure[]{
  \includegraphics[width=3.5cm]{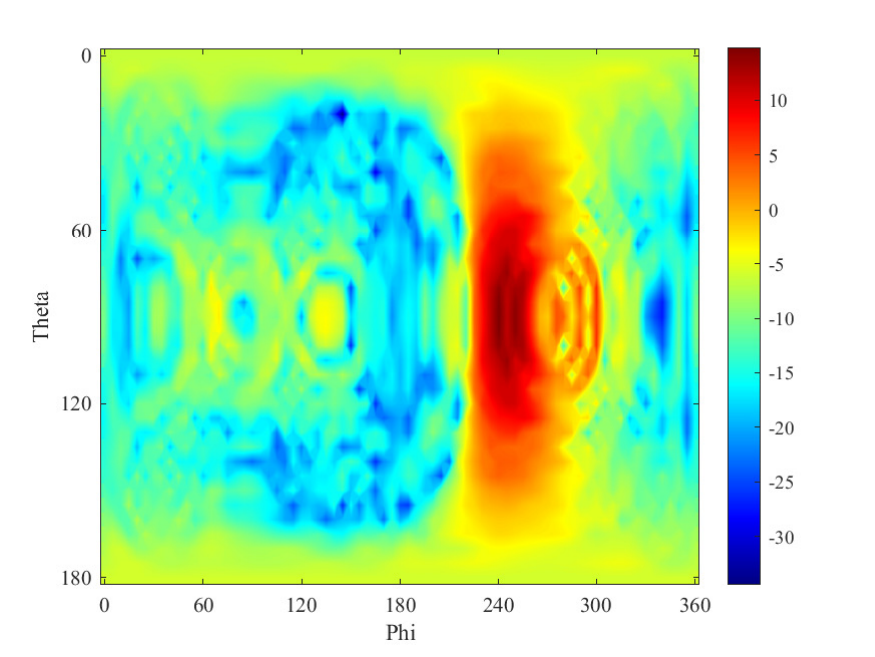}
  }

  \caption{Field distribution around the waveguide-PA interface and 2D antenna gain heat maps. (a), (b) Square PA. (c), (d) Triangular PA.}\label{field.Fig}
\end{figure}

As shown in Fig. \ref{simu_model.Fig}, a square PA block with 12 mm length and 3 mm thickness is placed in direct contact with the dielectric waveguide. The excitation is applied at Port 1 (P1) in Fig. \ref{simu_model.Fig}, indicating that the signal propagates from P1 to P2. As illustrated in Fig. \ref{field.Fig}(a), the electric field is confined within the dielectric waveguide and is locally coupled out at the region where the PA makes contact with the waveguide. In addition, Fig. \ref{field.Fig}(b) presents the antenna gain of the square PA, which reflects the radiation gain produced by the square PA in free space. As shown in Fig. \ref{field.Fig}(b), the max antenna gain is at $\phi = 252^\circ$ with 14.90 dB. The main lobe spans approximately $\phi \in [226^\circ, 256^\circ]$, and several pronounced side lobes are observed in $\phi \in [108^\circ, 130^\circ]$.





\subsection{Regular Triangular Pinching Antenna}
Similar to Fig. \ref{simu_model.Fig}, a regular triangular PA with side length of $12\sqrt{3}\approx 20.78 $ mm and thickness of 3 mm is placed in direct contact with the dielectric waveguide. As indicated in Fig. \ref{field.Fig}(c), the triangular PA effectively couples the electric field out of the waveguide. Fig. \ref{field.Fig}(d) depicts the antenna gain of the regular triangular PA, with a peak of 14.80 dB at $\phi = 240^\circ$. The main lobe spans approximately $\phi \in [234^\circ, 256^\circ]$. From Fig. \ref{field.Fig}(d), the main lobe gain can be 10 dB higher than the side lobes, indicating the directivity for wireless links.


The above results show that PA geometry is a key design parameter, because it directly changes the induced polarization-current distribution and, in turn, determines the radiated field and the resulting free-space antenna gain. In particular, the simulation results indicate that a square dielectric block generates a relatively broad main lobe toward the signal transmission direction and produces clear side lobes, whereas a regular triangular block provides stronger directivity with a narrower main lobe and weaker side lobes. 

With a simple geometry, the PAs behave as directive radiators. This finding suggests that near-isotropic radiation requires careful geometric design, after which the PAs can be modeled as isotropic radiators in the system analysis. This trend also agrees with the above discussed theory, where the geometry of the PA can be adjusted to reshape the phase distribution of the induced currents, so the radiation directivity can be achieved by hardware rather than relying only on conventional beamforming or PA placement algorithms.

\subsection{Arc Pinching Antenna}
From the square and triangular cases, we observe that modifying boundary geometry redistributes the field distribution and leads to a noticeable change in radiation directionality. This motivates us to seek a geometry with a continuous tuning knob for pattern steering.

The far-field pattern is determined by the spatial distribution of the induced polarization current. By modifying the PA boundary, the propagation path lengths inside the dielectric are changed, which reshapes the amplitude and phase distribution of the equivalent radiating aperture and deflects the far-field main lobe direction. Then, the arc PA is proposed, which can adjust the amplitude centroid in a controlled and efficient way.

\begin{figure}[!t]
  \centering
  \subfigure[]{
  \includegraphics[width=3.5cm]{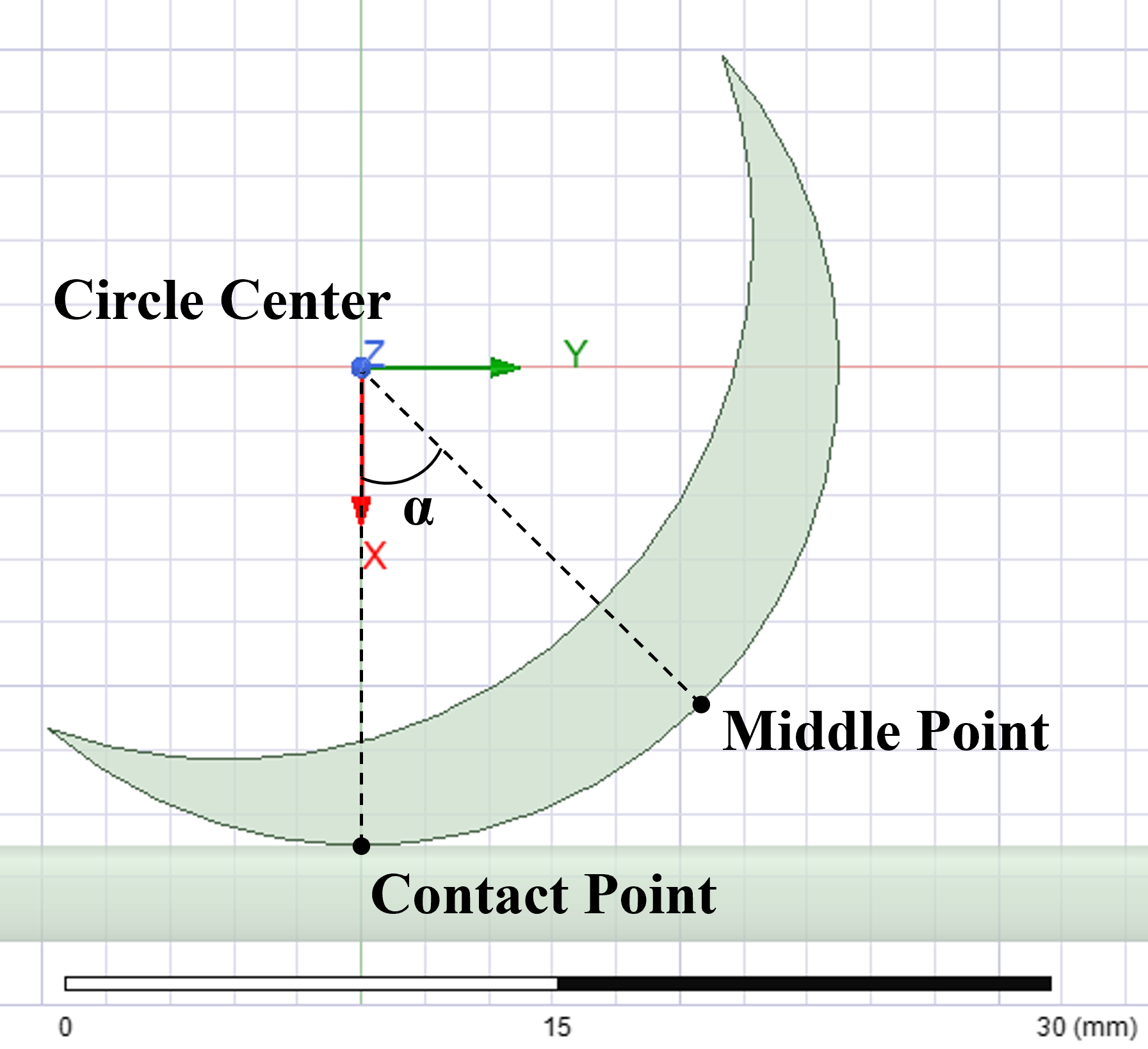}
  }
  \quad
  \subfigure[]{
  \includegraphics[width=3.5cm]{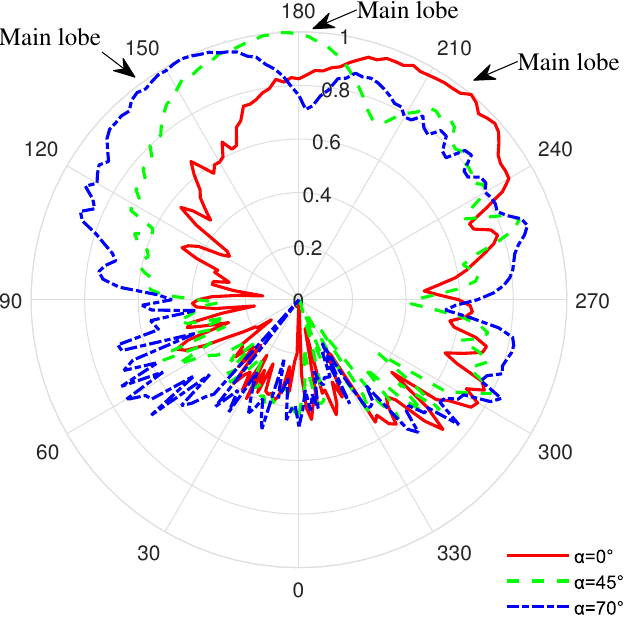}
  }
  
  \caption{Structure and radiation pattern of the proposed arc PA. (a) The arc PA. (b) Radiation patterns of arc PA with different rotation angles.}\label{arcPA.Fig}
\end{figure}

As shown in Fig. \ref{arcPA.Fig}(a), the proposed arc PA has a radius of 15 mm and thickness of 3 mm. Rotating the arc PA changes the rotation angle $\alpha$, which effectively alters the excitation location and consequently reshapes the radiation pattern of the PA system. Fig. \ref{arcPA.Fig}(b) presents the normalized simulated radiation patterns for different $\alpha$. As $\alpha$ varies, the main lobe direction steers from $150^\circ$ to $210^\circ$, demonstrating that the arc PA provides a simple geometric degree of freedom for controlling the radiation direction.

In practice, geometry-based directivity can reduce the need for dense PA activation and complex position-search procedures when the intended coverage direction is known, such as corridors or fixed hotspots, offering a low-cost way to steer energy toward a target direction. 

\section{Prototype and Experimental Results}\label{section:measurement}
To validate the simulation results and assess the practical feasibility of the proposed PA system, this section describes the developed prototype and the corresponding measurement setup. Radiation-pattern measurements for square and triangular PAs were performed, and the measured radiation characteristics were compared against the simulation results.

\subsection{Prototype}
\begin{figure} [t!]
  \centering
  \includegraphics[width= 3.12in, height=0.71in]{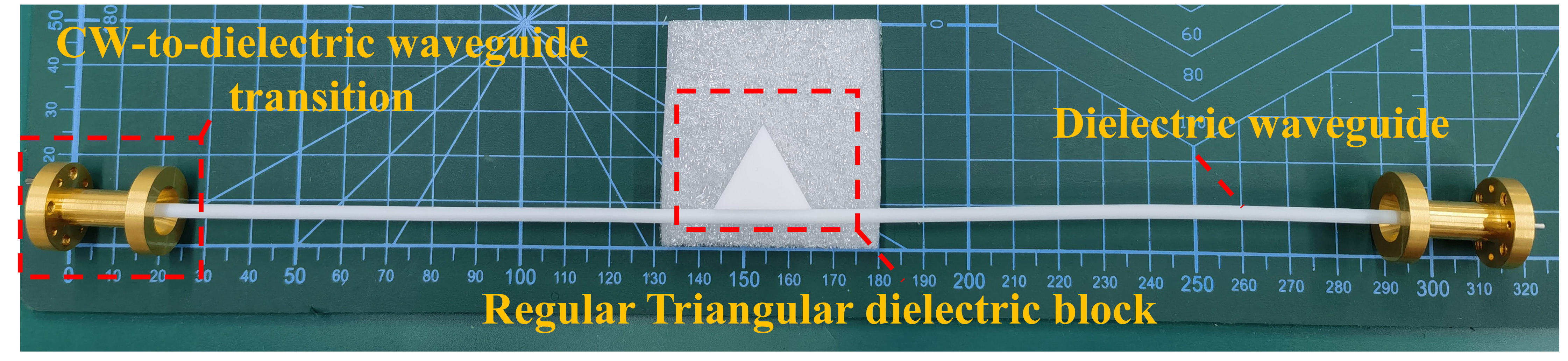}
  \caption{The prototype of the proposed PA system with a regular triangular dielectric block as the PA.}\label{prototype.Fig}
\end{figure}

Fig. \ref{prototype.Fig} shows the prototype of the proposed PA system. At the center frequency of 60 GHz, the RF signal is fed into the dielectric waveguide through one transition, while the other transition is terminated. The dielectric block (i.e., the PA) is placed in close proximity to the waveguide to couple energy into it and generate radiation to free space, thereby establishing wireless transmission links. Since the PA element can be readily installed or removed, the proposed architecture provides a convenient and reconfigurable approach for signal transmission.

\subsection{Measurement Deployment}
\begin{figure}[!t]
  \centering
  \subfigure[]{
  \includegraphics[width=3.5cm]{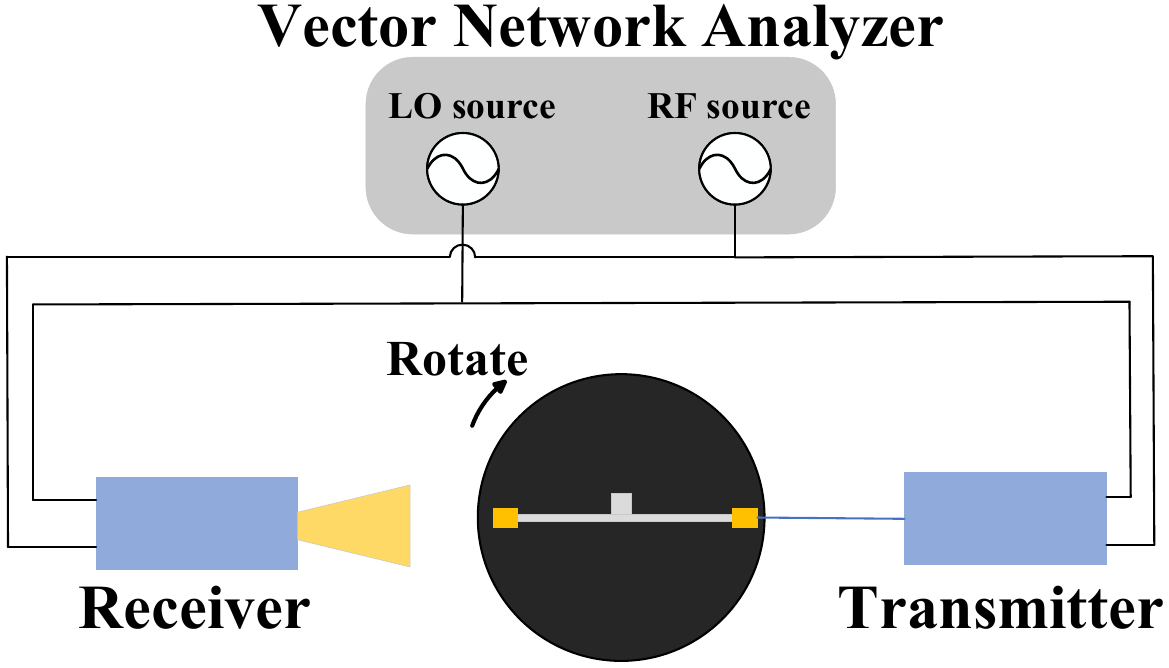}
  }
  \quad
  \subfigure[]{
  \includegraphics[width=3.5cm]{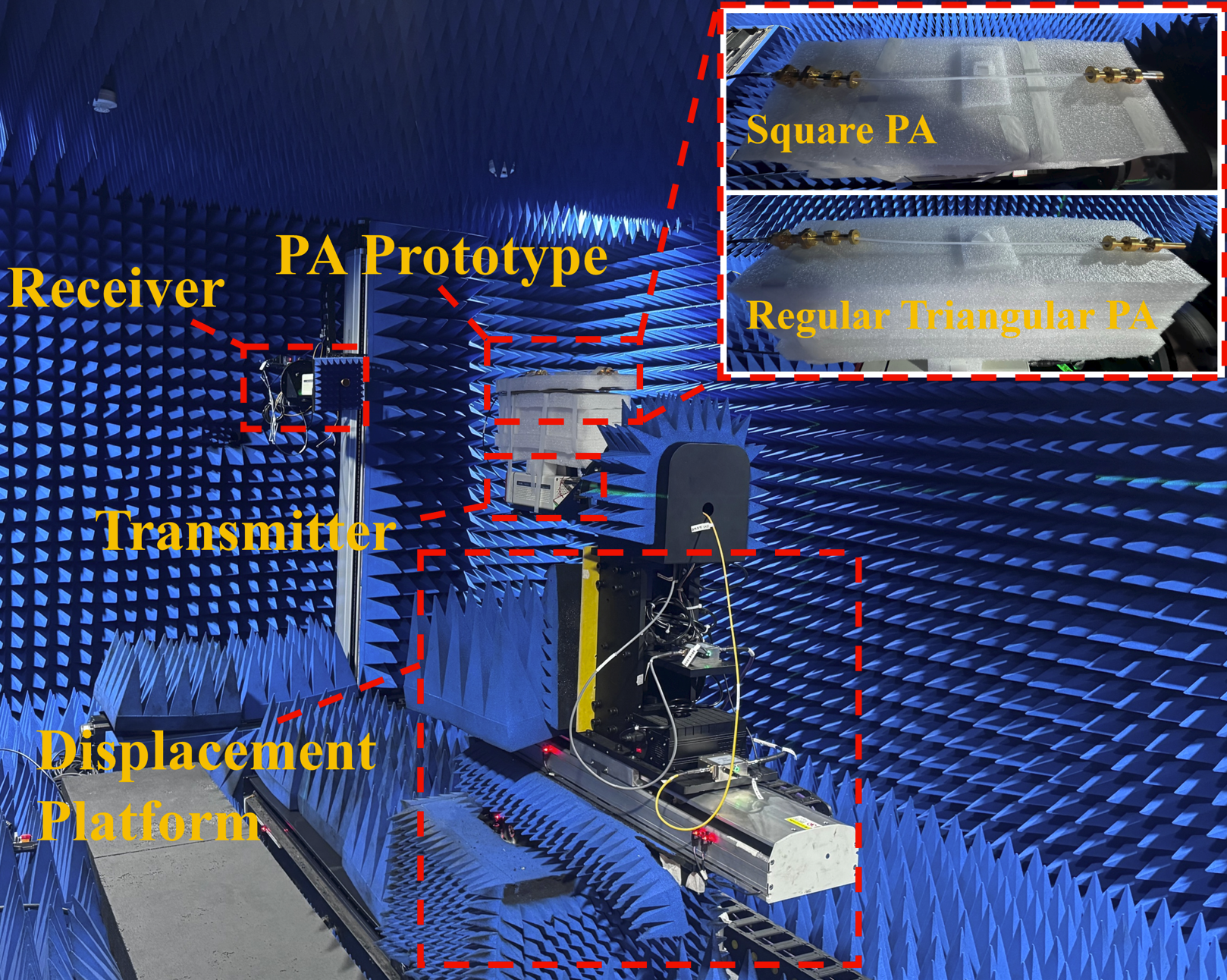}
  }
  
  \caption{Prototype measurements for the proposed PA systems. (a) The architecture of the measuring system with the proposed PA system. (b) Overview of the radiation pattern measurement platform for the proposed PA systems.}\label{measure.Fig}
\end{figure}

As shown in Fig. \ref{measure.Fig}(a), the measurement system for PA radiation pattern consists of transmitter and receiver modules, a computer-controlled rotary displacement platform and a vector network analyzer (VNA). The VNA generates RF and provides local oscillator (LO) sources. The transmitter radiates the RF signal through the PA, and the receiver captures it using a horn antenna.

The proposed PA prototype was mounted on the displacement platform that enables $360^\circ$ rotation. The horn antenna at the receiver was positioned in the same plane as the PA to measure the azimuth-plane gain pattern corresponding to the $\theta = 90^\circ $ cut in antenna gain heat map of PAs. To eliminate the influence of the system hardware, calibrations were conducted before measurements. All measurements were conducted at 60 GHz, and the distance between the horn antenna and the PA was 1.6 m. During the measurements, the rotary displacement platform swept at a constant rate of $1^\circ$/s. The measurement system is illustrated in Fig. \ref{measure.Fig}(b), with the key hardware modules highlighted by dashed rectangular boxes.



\subsection{Experimental Results}
\begin{figure}[!t]
  \centering
  \subfigure[]{
  \includegraphics[width=3.5cm]{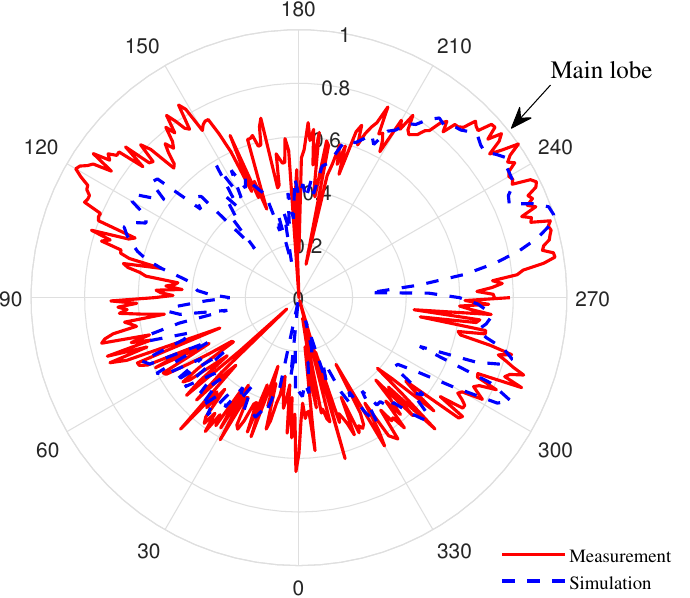}
  }
  \quad
  \subfigure[]{
  \includegraphics[width=3.5cm]{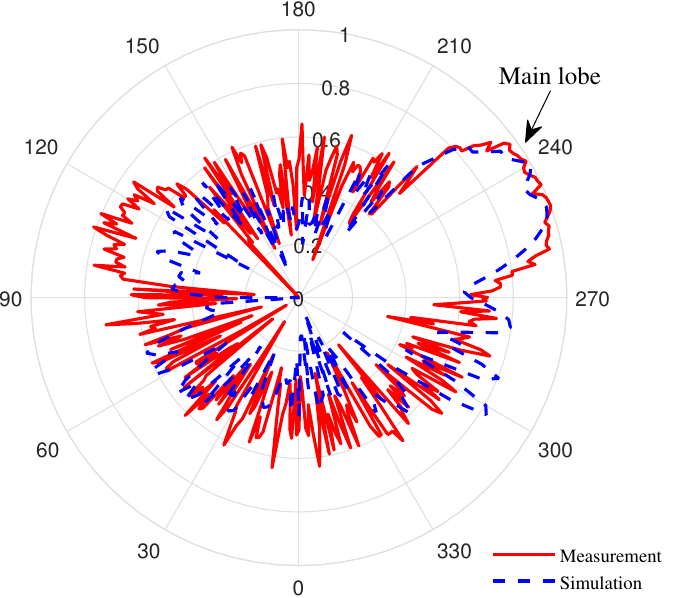}
  }
  
  \caption{Radiation pattern of the proposed PA system. (a) Square PA. (b) Triangular PA.}\label{2Dpattern.Fig}
\end{figure}

Fig. \ref{2Dpattern.Fig}(a) compares the normalized measured and simulated azimuth-plane radiation patterns for the $\theta=90^\circ$ cut in Fig. \ref{field.Fig}(b). The measurement agrees well with the simulation in the main lobe, exhibiting a similar beamwidth and propagation direction. Moreover, several pronounced side lobes are observed between $120^\circ$ and $150^\circ$, indicating multi-beam radiation in the azimuth plane. These results confirm that the proposed PA system with a square PA can radiate the signal to multiple directions, which can be leveraged to serve multiple wireless users effectively.


Fig. \ref{2Dpattern.Fig}(b) presents the normalized measured and simulated of azimuth-plane radiation patterns for the cut at $\theta = 90^\circ$ in Fig. \ref{field.Fig}(d). Both results exhibit a dominant main lobe centered near $240^\circ$, with close agreement in beamwidth. The measurements show a dominant main lobe, with clearly suppressed side lobes in the normalized radiation patterns. Overall, the results verify that the proposed PA system employing a regular triangular PA provides strong directivity.


In conclusion, the main lobe direction agrees well between the simulation and the measurement. The small differences and ripples outside the main lobe are caused by measurement and fabrication tolerances. The measurements confirm the simulation results and show that PA geometries strongly affect the radiation pattern. Compared with the square PA, the triangular PA provides higher directivity, while the square PA offers multiple beams that can serve users in more directions. 

These results highlight the importance of PA geometry design in PA systems. From \eqref{pinch_channel.eq}, \eqref{received_signal.eq} and \eqref{transmit_signal.eq}, the received signals at wireless users can be affected by the location of PAs as the PAs positions determine the phase shifts of the wireless signals. By properly adjusting the locations of PAs, the received signals can be constructively combined at users \cite{xu2025rate}. However, even with correct phase alignment, the achievable gain is limited if the user lies in a weak side lobe region. Moreover, the results indicate that PA geometry itself provides an additional degree of freedom, since shaping the geometry can steer the radiation direction without changing the PA locations. Therefore, a suitable PA geometry is required not only to fully exploit the existing location-tuning-based beamforming but also to enable direct direction control at the hardware level. 

\vspace{-0.1cm}
\section{Conclusion}\label{section:con}
This letter investigated the geometry-dependent radiation behavior of PA systems and demonstrated flexible pattern shaping through structural reconfiguration. The simulations and prototype measurements were performed to characterize the radiation behavior of square and regular triangular PA. The results showed clear differences in the resulting patterns for the square and triangular designs, which highlighted the key role of PA geometry in PA-assisted wireless communication systems. In addition, an arc-shaped PA is proposed as a simple and low-cost approach to adjust the radiation direction. 


\bibliographystyle{IEEEtran}
\bibliography{refPinchAntenna}

\end{document}